\documentclass[aip,preprint]{revtex4-1}

\usepackage{graphicx}

\draft 

\begin{document}


\title{Non-conventional graphene superlattices as electron band-pass filters} 



\author{A. S\'anchez-Arellano}
\author{J. Madrigal-Melchor}
\author{I. Rodr\'iguez-Vargas}
\affiliation{Unidad Acad\'emica de F\'isica, Universidad Aut\'onoma de Zacatecas, Calzada Solidaridad Esquina con Paseo La Bufa S/N, 98060 Zacatecas, Zac., Mexico.}


\date{\today}

\begin{abstract}
Electron transmission through different gated and gapped graphene superlattices (GSLs) is studied. Linear, Gaussian, Lorentzian and P\"oschl-Teller superlattice potential profiles have been assessed. A relativistic description of electrons in graphene as well as the transfer matrix method have been used to obtain the transmission properties. We find that is not possible to have perfect or nearly perfect pass bands in gated GSLs. Regardless of the potential profile and the number of barriers there are remanent oscillations in the transmission bands. On the contrary, nearly perfect pass bands are obtained for gapped GSLs. The Gaussian profile is the best option when the number of barriers is reduced, and there is practically no difference among the profiles for large number of barriers. We also find that both gated and gapped GSLs can work as omnidirectional band-pass filters. In the case of gated Gaussian GSLs the omnidirectional range goes from -$50^{\circ}$ to $50^{\circ}$ with an energy bandwidth of 55 meV, while for gapped Gaussian GSLs the range goes from -$80^{\circ}$ to $80^{\circ}$ with a bandwidth of 40 meV. Here, it is important that the energy range does not include remanent oscillations. On the light of these results, the hole states inside the barriers of gated GSLs are not beneficial for band-pass filtering. So,  the flatness of the pass bands is determined by the superlattice potential profile and the chiral nature of the charge carriers in graphene. Moreover, the width and the number of electron pass bands can be modulated through the superlattice structural parameters. We consider that our findings can be useful to design electron filters based on non-conventional GSLs.
\end{abstract}

\pacs{}

\maketitle 

\section{Introduction}
Semiconductor superlattices are essential as injector and/or active region in several applications. In quantum cascade lasers,\cite{faist1994,scamarcio1997} superlattices with well-defined stop (high reflection) and pass (high transmission) bands are needed for good device efficiency. Since its seminal proposal,\cite{tung1996,gomez1999} semiconductor superlattices with Gaussian potential profile are regarded as the archetypal structure to obtain flat stop and pass bands with nearly 100\% reflection and transmission probability. The fundamental properties of these non-conventional superlattices were experimentally verified.\cite{diez2000} The electron-electron interaction and disorder effects were also studied,\cite{diez2000,banfi2001} having a limited influence on the filtering characteristics. The striking features of the so-called Gaussian superlattices sparked a lot of interest, deriving in multiple studies.\cite{diez2000,banfi2001,yang2000,gu2000,pacher2001,morozov2002,sprung2002,sprung2003,martorell2004,gomez2004,xie2007} These studies have been extended to other elemental excitations.\cite{liu2006,tenorio2009,villegas2005,li2011p,li2011,madrigalmelchor2017} Recently, non-conventional potential profiles have been used to improve the efficiency of thermoelectric devices and magnetic tunnelling junctions.\cite{karbaschi2016,priyadarshi2018,sharma2018} The common factor in all these studies is the use of the Gaussian profile or alike profiles to create nearly perfect stop and pass bands. 

Under this context, a possible electronics based on cutting-edge materials like graphene will need efficient devices that act as injector, collector and/or active region. In fact, the so-called gated GSLs,\cite{park2008,barbier2010,garciacervantes2016} a graphene sheet with a periodic arrangement of top gates, represent a possibility. In these superlattices the propagation of charge carriers is highly anisotropic.\cite{park2008,barbier2010} The energy minibands and minigaps depend strongly on the angle of the impinging electrons, opening the possibility for an angle-dependent bandgap engineering.\cite{garciacervantes2016} The transmission minigaps (stop bands) can be tuned from meV to eV by changing the angle of incidence. An alternative is to fix the angle of incidence and enlarge the stop bands by using two or more gated GSLs with different periodic potentials.\cite{cheng2011} To get the stop-band enlargement is fundamental that the bandgaps of the constituent superlattices overlap. Another possibility are aperiodic and non-conventional gapped graphene heterostructures.\cite{zhang2012,zhang2014,sattari2018}  In this case, the  omnidirectional electronic bandgap associated to gapped graphene can be extended  by appropriately choosing the widths of the potentials of the constituent superlattices.\cite{zhang2012,zhang2014} It is also possible to have a tunable band-stop filter by modulating the Fermi veolicty barriers and the external bias voltage in non-conventional fashion.\cite{sattari2018} 

As we have corroborated much attention has been paid to create and modulate stop bands in graphene superlattices. However, for superlattice devices pass bands are as important as stop bands. Unfortunately, the literature about pass bands in GSLs, specifically about how flattening them,\cite{lu2015} is scare. For instance, magnetic GSLs with Gaussian modulation in the heights of the magnetic barriers could be an option for band-pass filters.\cite{lu2015} However, magnetic modulation with precise potential profiles could be tricky. Actually, gating is a more natural technique for 2D materials. In fact, most of the exotic and unprecedented phenomena in graphene have been tested with the help of this technique. With gating we will have electrostatic barriers with hole states inside them. These states are fundamental for most of the exotic phenomena in graphene, like Klein tunnelling.\cite{katsnelson2006,young2009} However, as far as we know, the role of them in band-pass filtering has not been addressed. Other issue that has been controversial since the beginning of non-conventional superlattices is the shape of the potential profile.\cite{tung1996,yang2000,sprung2003,zhang2014} Some works claim that Gaussian superlattices are the best for band-pass filtering,\cite{tung1996} others say that there is nothing special with the Gaussian profile and that any potential with smooth variation can serve as good band-pass filter.\cite{yang2000,zhang2014} 

The aim of the present work is twofold: first, we want to know the relevance of hole states for band-pass filtering, in order to do 	so, we will compare the transmission properties of two antagonistic systems, gated and gapped non-conventional GSLs; second, we want to find out to what extent the shape of the potential profile is preponderant for the filtering characteristics, so we will consider different non-conventional GSLs such as Gaussian, Lorentzian, Linear and P\"oschl-Teller. 

\section{Model and Method}
\begin{figure}
 \includegraphics[scale=0.4]{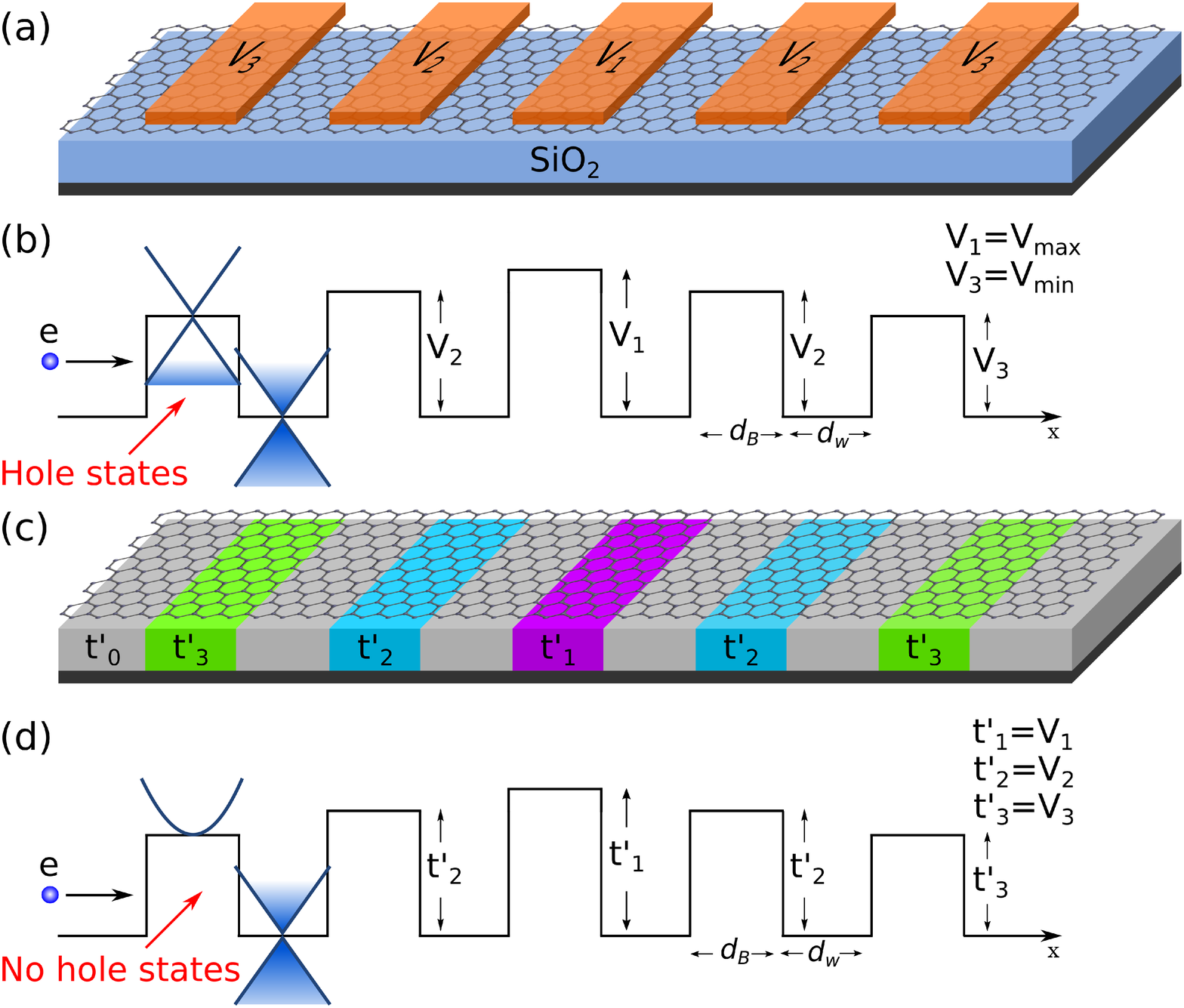}
 \caption{\label{Fig1} (a) Schematic representation (top view) of the possible gated GSLs. The orange stripes represent metallic electrodes at different potential energies. (b) Resulting potential profile of (a). Here, $V_1$ and $V_3$ represent the maximum ($V_{max}$) and minimum ($V_{min}$) potential barriers in the system. $d_B$ and $d_W$ are the widths of the barrier and well regions. (c) and (d) the same as in (a) and (b), but for gapped GSLs. In this case the potential barriers are generated by substrates with different degree of interaction with the graphene sheet ($t'_i$). The main difference between gated and gapped GSLs is that in the former there are hole states inside the barriers. By appropriately choosing the heights of the potential barriers we can obtain superlattices with Linear, Gaussian, Lorentzian and P\"oschl-Teller potential profiles, see Fig. 2. }%
 \end{figure}

\begin{figure}
 \includegraphics[scale=0.5]{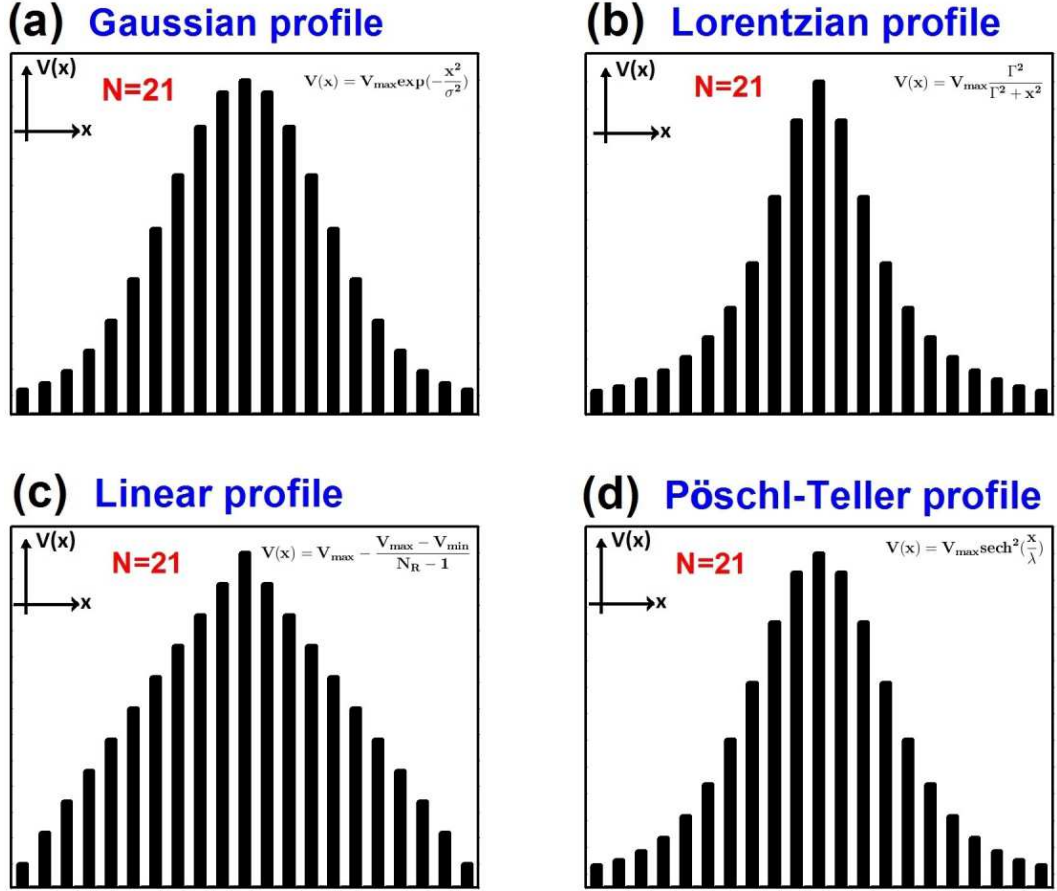}
 \caption{\label{Fig2} (a) Gaussian, (b) Lorentzian, (c) Linear and (d) P\"oschl-Teller potential profiles of gated and gapped GSLs. In all cases the number of barriers is $N=21$. The height of the potential varies in Gaussian, Lorentzian, Linear and P\"oschl-Teller fashion going from a minimum value at the edges of the structure to a maximum value at the center of it. The mathematical function used in each case is also represented. Here, the function parameters ($\sigma$, $\Gamma$, $N_R$ and $\lambda$) are adjusted to the total length of the superlattice such that the minimum and maximum barrier heights be the same for all profiles.}%
 \end{figure}

The schematic representation of the possible devices for non-conventional gated and gapped graphene superlattices is shown in Fig. \ref{Fig1}. In the case of gated GSLs (Fig. \ref{Fig1}a) the graphene sheet is placed on a non-interacting substrate like SiO$_2$ and top gates are patterned to induce the non-conventional potential profile (Fig. \ref{Fig1}b). The device also includes, not shown, a back gate and left and right leads. We can have potential barriers of different height by modulating the strength of the applied electrostatic field. It is also important to remark that there are hole states inside the barriers because the main effect of gating is a shifting of the Dirac cones, see Fig. \ref{Fig1}b. In the case of gapped GSLs (Fig. \ref{Fig1}c) the potential barriers are induced by substrates with different degree of interaction with the graphene sheet. This interaction results in a bandgap and no hole states inside the potential barriers, see Fig. \ref{Fig1}d. So, by gating or interacting substrates, in principle, is possible to have non-conventional gated and gapped GSLs, respectively. In our case, we will consider Gaussian, Lorentzian, Linear and P\"oschl-Teller potential profiles, see Fig. \ref{Fig2}. 

The semi-infinite and quantum well regions, in both gated and gapped GSLs, are described by the well-known monolayer graphene Hamiltonian $H_{free}=\hbar v_F {\bf \sigma} \cdot {\bf k}$, with linear dispersion relation $E=\pm \hbar v_F k$ and eigenfunctions, 

\begin{equation}
\psi_{k}^{\pm} (x,y) = \frac{1}{\sqrt{2}}
\left(\begin{array}{c}
1 \\
u_{\pm}
\end{array}\right)
e^{\pm ik_{x}x + ik_{y}y},
\label{WFk}
\end{equation} 

\noindent where $v_F$ is the Fermi velocity, $k$ is the magnitude of the wave vector in these regions, $k_x$ and $k_y$ are
the longitudinal and transverse components of $k$, and $u_{\pm} = \pm \mbox{sgn}(E) e^{\pm i\theta}$ the coefficients of the wave
functions that depend on the angle of the impinging electrons, $\theta = \arctan(k_y/k_x)$. The Hamiltonian for the gated regions is $H_{gated}=\hbar v_F {\bf \sigma} \cdot {\bf q_i}+ V_i {\bf 1}$, with dispersion relation $E - V_i=\pm \hbar v_F q_i$ and eigenfunctions, 

\begin{equation}
\psi_{q_i}^{\pm} (x,y)= \frac{1}{\sqrt{2}}
\left(\begin{array}{c}
1 \\
v_{\pm,i}
\end{array}\right)
e^{\pm iq_{x,i}x + iq_{y,i}y},
\label{WFq}
\end{equation} 

\noindent here $V_i$ represents the strength of the electrostatic potential of the $i$-th barrier, $q_i$ is the corresponding wave vector, $q_{x,i}$ and $q_{y,i}$ are the components of $q_i$, and $v_{\pm,i}$ the coefficients of the wave functions.\cite{vargas2012} For the gapped regions  $H_{gapped}=\hbar v_F {\bf \sigma} \cdot {\bf q_i}+ t'_i \sigma_z$ and $E^2-t_i'^2=\hbar^2 v_F^2 q_i^2$. The eigenfunctions have the same mathematical form as Eq. (\ref{WFq}), however $q_{x,i}$ and $v_{\pm,i}$ depend on the bandgap energy $t'_i=E_{g,i}/2$.\cite{vargas2012}

This information, eigenfunctions and wave vectors, is essential to compute the transfer matrix of the system, and with it the transmission probability or transmittance. By imposing the continuity condition for the wave function along the superlattice structure as well as the conservation of the transverse momentum, we can obtain the transfer matrix of the system. For a non-conventional GSL of three barriers ($N=3$) the transfer matrix can be written as 

\begin{equation}
M=M_3 M_W M_2 M_W M_1 M_W M_2 M_W M_3,
\end{equation}

\noindent where $M_W$ and $M_i$ are the transfer matrices of the well and barrier ($i$-th) regions. These matrices depend on the so-called dynamic and propagation matrices.\cite{vargas2012} The energy and angle dependent transmittance can be obtained through the (1,1) element of the transfer matrix

\begin{equation}
T(E,\theta)=\frac{1}{|M_{11}|^2}.
\end{equation}

\section{Results and Discussion}
\begin{figure}
 \includegraphics[scale=0.5]{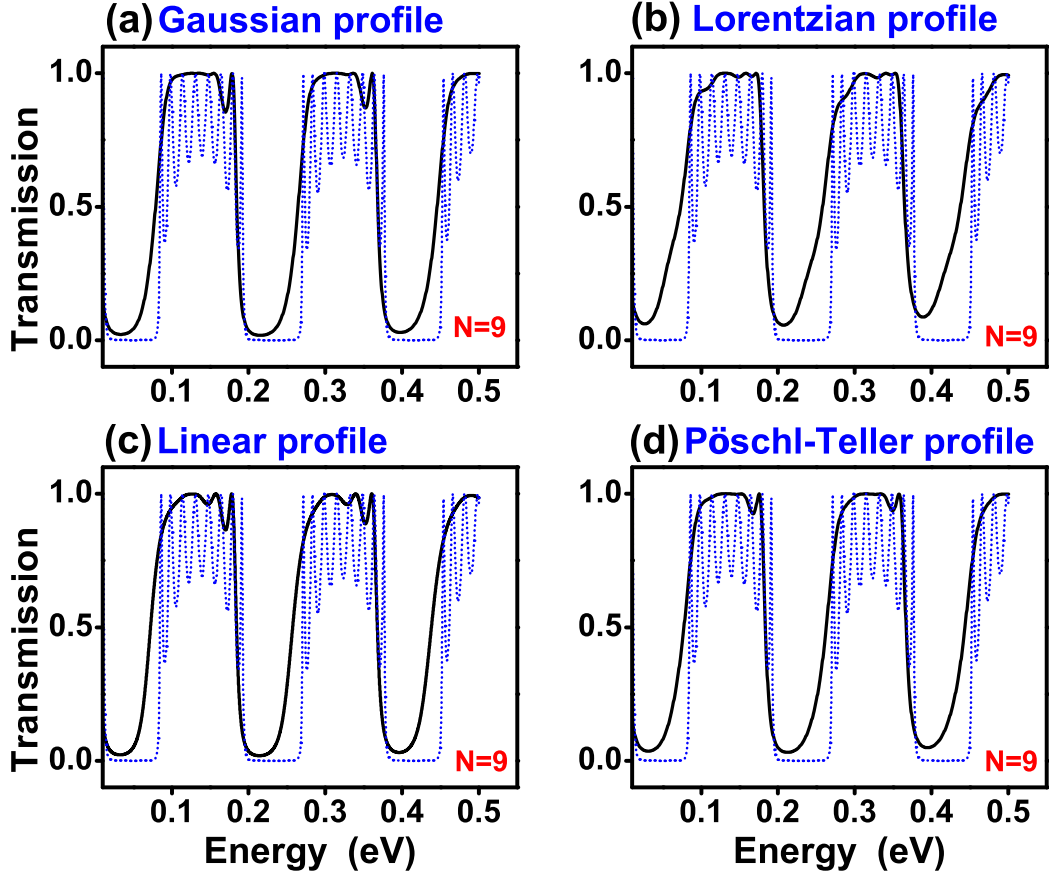}
 \caption{\label{Fig3} Transmission probability or transmittance versus the energy of the incident electrons for gated graphene superlattices with (a) Gaussian, (b) Lorentzian, (c) Linear and (d) P\"oschl-Teller potential profiles. In all cases the number of barriers, the heights of the maximum and minimum potential barriers, the widths of the barrier and well regions as well as the angle of incidence were the same. The specific values for all these quantities were $N=9$, $V_{max}=0.13$ eV, $V_{min}=0.01$ eV, $d_B=20a$, $d_W=80a$ and $\theta=45^{\circ}$, respectively. Here, $a$ represents the carbon-carbon distance in graphene, $0.142$ nm.}%
 \end{figure}

\begin{figure}
 \includegraphics[scale=0.5]{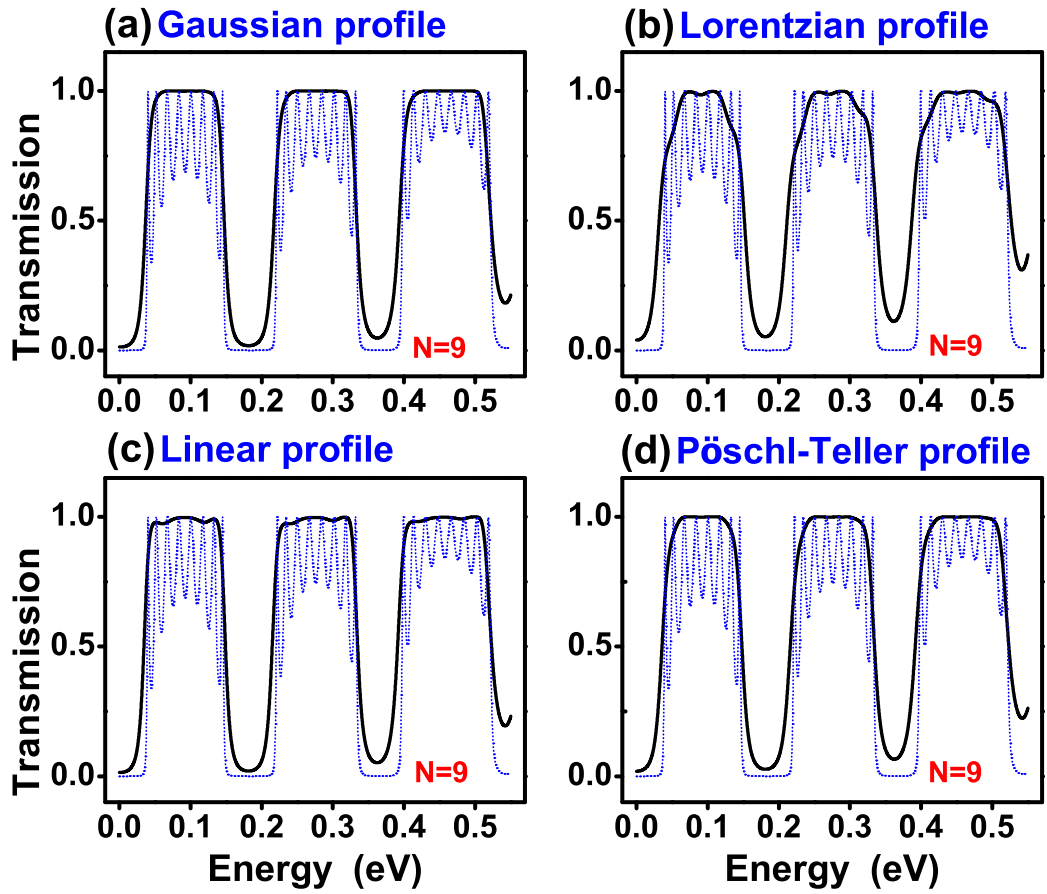}
 \caption{\label{Fig4} The same as in Fig. \ref{Fig3}, but for gapped GSLs.}%
 \end{figure}

\begin{figure}
 \includegraphics[scale=0.5]{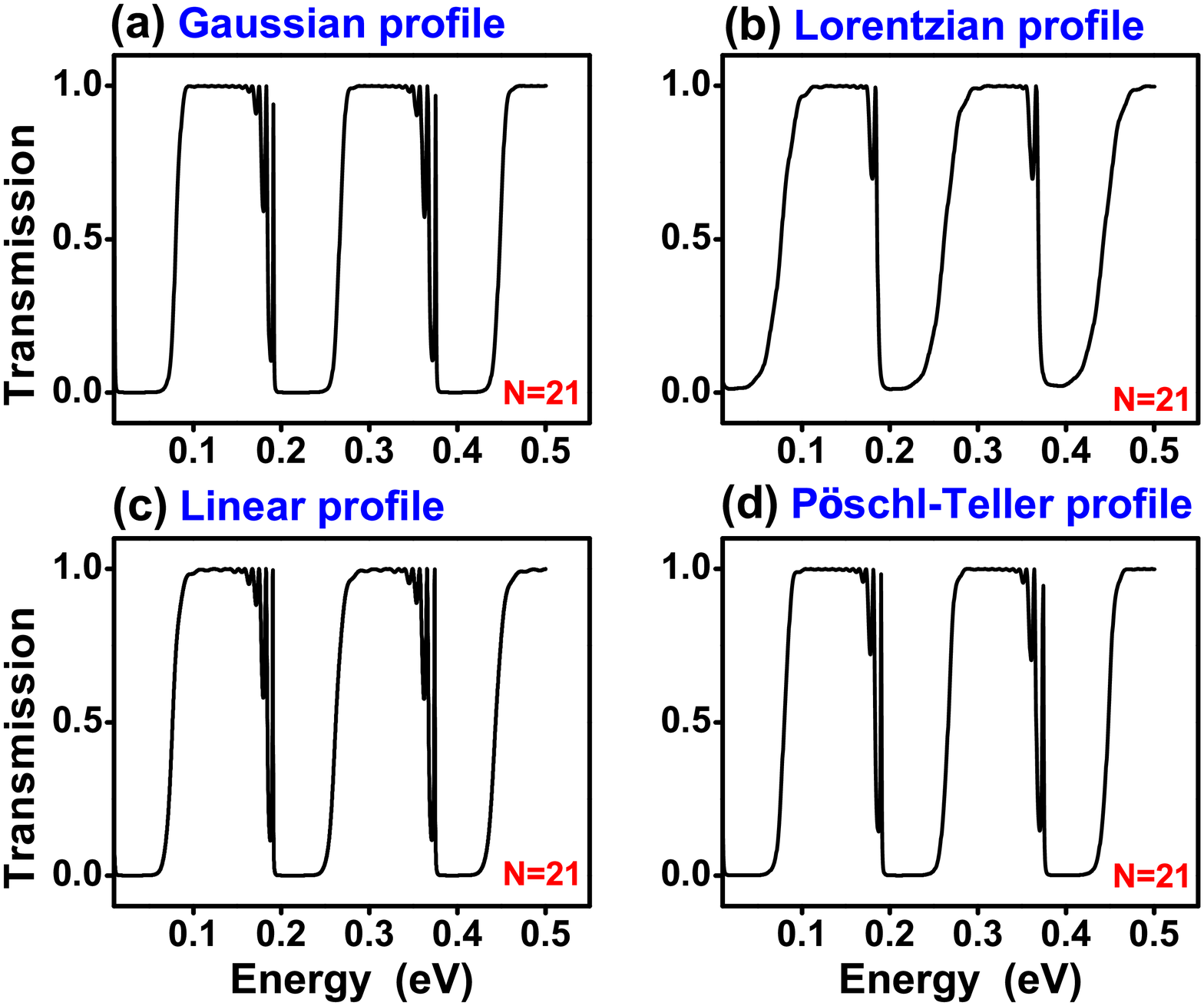}
 \caption{\label{Fig5} The same as in Fig. \ref{Fig3}, but for $N=21$.}%
 \end{figure}

\begin{figure}
 \includegraphics[scale=0.5]{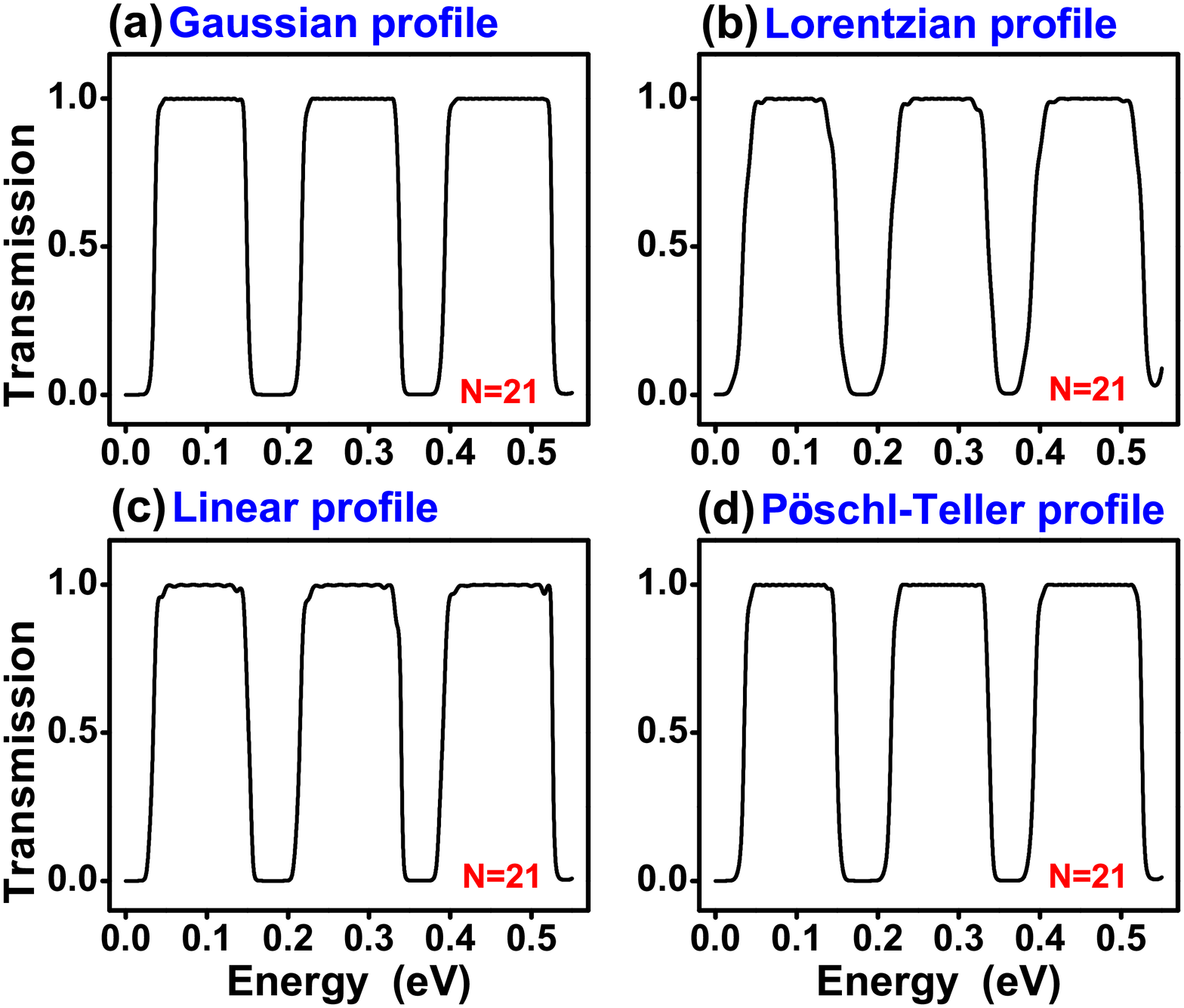}
 \caption{\label{Fig6} The same as in Fig. \ref{Fig4}, but for $N=21$.}%
 \end{figure}

In Fig. \ref{Fig3} we show the transmittance of (a) Gaussian, (b) Lorentzian, (c) Linear and (d) P\"oschl-Teller gated GSLs. We consider the same structural parameters for all non-conventional gated GSLs. In concrete, the number of barriers, the maximum and minimum height of barriers, the width of barriers and wells as well as the angle of incidence considered were: $N=9$, $V_{max}=0.13$ eV, $V_{min}=0.01$ eV, $d_B=20a$, $d_W=80a$ and $\theta=45^{\circ}$, respectively. The widths are given in terms of the carbon-carbon distance in graphene $a=0.142$ nm. We have included the transmittance of a uniform gated GSL, dotted-blue curves, as reference. As we can notice practically all non-conventional profiles reduce notably the transmittance oscillations of the uniform superlattice. We can also see that, regardless of the potential profile, the pass bands present a high probability, close to 100 \%. However, the pass bands are not well-defined, they have a rounded shape. For Gaussian and P\"oschl-Teller GSLs a notch (small oscillation) arises at the high energy side of the pass bands. Linear GSLs have more than a notch (more oscillations) and Lorentzian GSLs have no oscillations, however the pass bands and stop bands are not as defined as in the other cases. In the case of gapped GSLs (Fig. \ref{Fig4}) the pass bands have practically no oscillations and are better defined for the Gaussian profile. In Fig. \ref{Fig5} we show the transmittance of gated GSLs for $N=21$. As we can see with the number of barriers the oscillations increase and become more pronounced as well as the stop and pass bands adopt a better rectangular form. The Lorentzian profile (Fig. \ref{Fig5}b) deserves a special mention because, despite that the stop and pass bands are not well-defined as in the other non-conventional GSLs, the remanent oscillations in the pass bands are substantially reduced, in both, number and intensity. In Fig. \ref{Fig6} we show the corresponding results for gapped GSLs. As we can notice the transmission characteristics are nearly the same for all non-conventional gapped GSLs. It is remarkable that the pass bands are essentially flat and with 100\% transmission probability. It is also important to highlight the perfect rectangular shape of the stop and pass bands of the Gaussian profile. If we further increase the number of barriers, results not shown, the remanent oscillations of gated GSLs increase even more, and the transmission characteristics, flat pass bands, among non-conventional gapped GSLs become equivalent. 
 
\begin{figure}
 \includegraphics[scale=0.5]{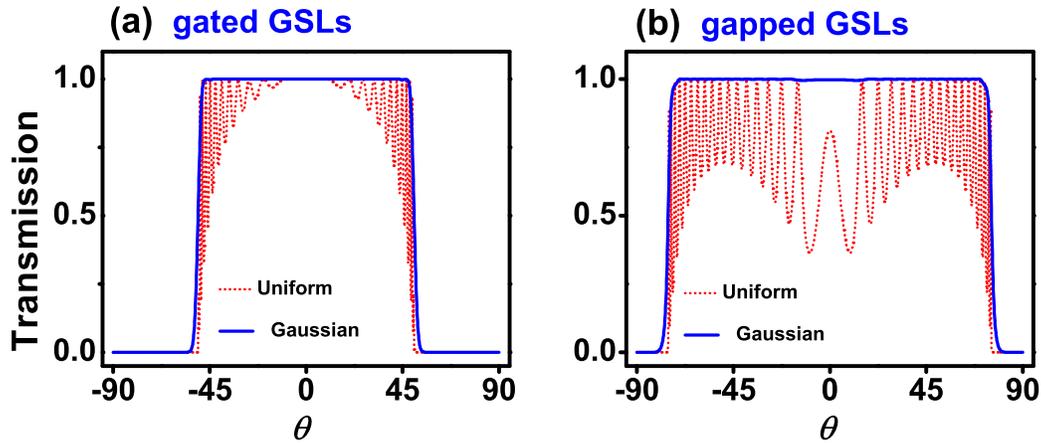}
 \caption{\label{Fig7} Angular distribution of the transmittance for (a) gated and (b) gapped Gaussian GSLs. The energy of the impinging electrons is $E=0.1$ eV. The transmittance for uniform GSLs (dotted-red curves) has been included as reference. The superlattice parameters are the same as in Figs. \ref{Fig5} and \ref{Fig6}.}%
 \end{figure}

\begin{figure}
 \includegraphics[scale=0.5]{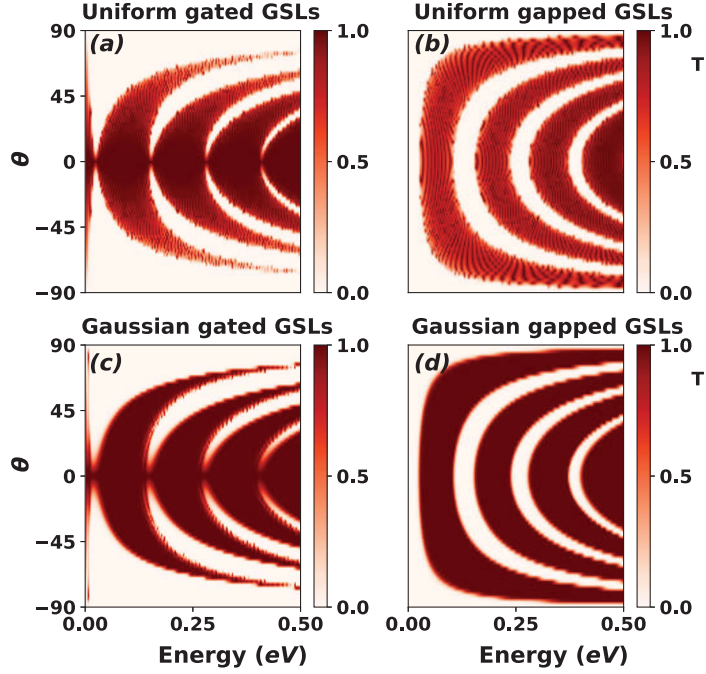}
 \caption{\label{Fig8} Contour maps $(E,\theta)$ of the transmittance for Uniform ((a) gated and (b) gapped) and Gaussian ((c) gated and (d) gapped) GSLs. The superlattice parameters are the same as in Fig. \ref{Fig7}.}%
 \end{figure}

\begin{figure}
\includegraphics[scale=0.5]{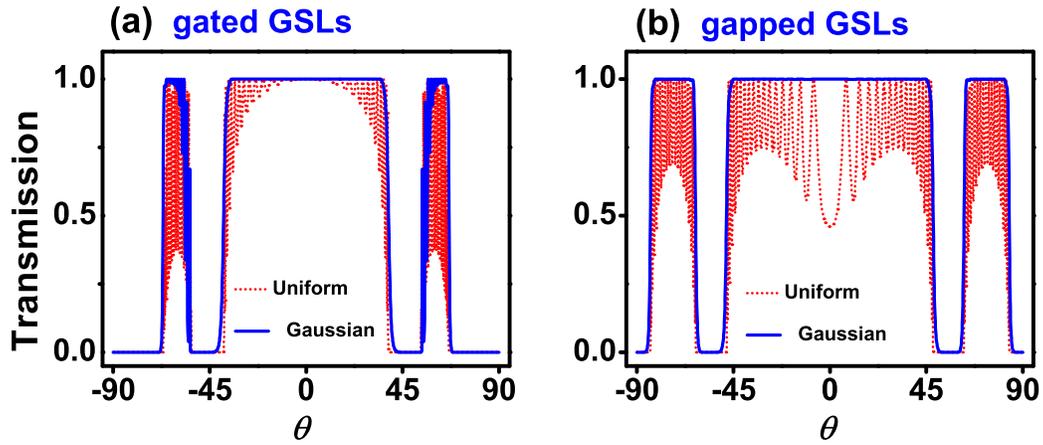}
 \caption{\label{Fig9} The same as in Fig. \ref{Fig7}, but for $E=0.23$ eV.}%
 \end{figure}

If we now fix the energy and vary the the angle of incidence 	we can obtain the so-called angular distribution of the transmittance $T(\theta)$. Fig. \ref{Fig7} shows $T(\theta)$ for (a) gated and (b) gapped Gaussian GSLs. We have included the corresponding results for uniform GSLs as reference. The energy of the impinging electrons was fixed to $E=0.1$ eV, which is lower than the maximum barrier height of the system. The superlattice parameters are the same as in Fig. \ref{Fig5} and \ref{Fig6}. As in the case of $T(E)$ the non-conventional profile eliminates the transmittance oscillations presented in uniform GSLs. In the gated case we can see a flat pass band that spans from -$50^{\circ}$ to $50^{\circ}$ with no oscillations at the edges. For gapped GSLs the range of perfect transmission is bigger, from -$80^{\circ}$ to $80^{\circ}$. These results are quite interesting because both gated and gapped non-conventional GSLs can work as omnidirectional filters. Here, it is fundamental to have control of the energy of the impinging electrons. In order to know what happen for other energies as well as to have a bigger picture of the transmission properties we have computed the contour maps $(E,\theta)$ of the transmittance. The corresponding results are shown in Fig. \ref{Fig8}. As we can notice the non-conventional profile gives rise to dispersionless transmission bands, uniform dark-red zones. We can also see that the omnidirectional pass bands of Fig. \ref{Fig7} have a considerable energy bandwidth. In particular, the bandwidth for gated GSLs is about 55 meV, while for gapped GSLs is about 40 meV. It is also important to mention that the high energy edge of the pass bands of non-conventional gated GSLs is not as uniform as the low energy edge. This non-uniformity is related to the remanent oscillations previously discussed. This point is quite relevant for gated GSLs because if we want to avoid the remanent oscillations we have to consider energies that do not include the high energy edge of the transmission bands. For instance, if we choose an energy of 0.23 meV, which is above the maximum barrier height, the outermost transmission bands of gated GSLs (Fig. \ref{Fig9}a) present remanent oscillations. On the contrary, gapped GSLs (Fig. \ref{Fig9}b) present perfect pass bands. Finally, we consider that some remarks are necessary: 1) We can attribute the remanent oscillations to the hole states inside the barriers of gated GSLs because this is the fundamental difference between gated and gapped GSLs. So, further analysis is needed in order to unveil why the oscillations are located in the high energy edge of the transmission bands as well as to design strategies to eliminate them; 2) In the last part of the present study we focus our attention to Gaussian GSLs due to the relevance of this profile, however similar results can be obtained for the other non-conventional profiles; 3) As in the case of superlattices of conventional materials the transmission characteristics of GSLs can be modulated through the superlattice parameters, results not shown; 4) Other 2D materials like bilayer graphene, silicene, phosphorene and transition-metal-dichalcogenides could be an option for electron band-pass filtering due to its intrinsic band structure characteristics. In particular, bilayer graphene represents an excellent option because a bandgap and the chirality of the charge carriers can be modulated with gating. 

\section{Conclusions}
In summary, we show that the chiral nature of the charge carriers in graphene as well as the superlattice potential profile are essential for good band-pass filtering. By comparing the transmission properties of gated and gapped GSLs of Gaussian, Lorentzian, Linear and P\"oschl-Teller potential profiles we obtain that the hole states inside the barriers of gated GSLs are the main obstacle for good band-pass filtering. Regardless of the potential profile and the number of barriers persistent oscillations in the transmission bands hamper the formation of perfect or nearly perfect pass bands. For gapped GSLs we obtain excellent band-pass filtering characteristics. Gaussian GSLs have the best filtering when the number of barriers is reduced, while all gapped GSLs are good filters, practically equivalent, for large number of barriers. Furthermore, we find that both gated and gapped non-conventional GSLs can work as omnidirectional band-pass filters. For gated GSLs is important that the omnidirectional energy bandwidth does not include remanent oscillations. We consider that our results can be useful in designing electron band-pass filters based on non-conventional GSLs. 

\bibliography{ASArellano2018}

\newpage

\end{document}